\documentclass[prb,a4paper,twocolumn,floatfix,showpacs,showkeys,amsmath,amssymb,nobibnotes,altaffilletter]{revtex4}

\usepackage{graphicx}
\usepackage{subfigure}
\usepackage{xspace}

\newcommand{\htwopc}{H$_2$Pc(OC14,10)$_4$\xspace}

\begin{document}

\preprint{}

\title{Charge Transport Properties of a Metal-free Phthalocyanine Discotic Liquid Crystal}

\author{C. Deibel}
\author{D. Janssen}
\author{P. Heremans}\email{heremans@imec.be}
\affiliation{IMEC, Kapeldreef 75, 3001 Leuven, Belgium}

\author{V. de Cupere}
\author{Y. Geerts}
\affiliation{Laboratoire de Chimie des Polym\`eres, CP 206/1, Universit\'e Libre de Bruxelles, Boulevard du Triomphe, 1050 Bruxelles, Belgium}

\author{M. L. Benkhedir}
\author{G. J. Adriaenssens}
\affiliation{Laboratorium voor Halfgeleiderfysica, University of Leuven, 3001 Leuven, Belgium}



\begin{abstract}
Discotic liquid crystals can self-align to form one-dimensional semiconducting wires, many tens of microns long. In this letter, we describe the preparation of semiconducting films where the stacking direction of the disc-like molecules is perpendicular to the substrate surface.  We present measurements of the charge carrier mobility, applying temperature-dependent time-of-flight transient photoconductivity, space-charge limited current measurements, and field-effect mobility measurements. We provide experimental verification of the highly anisotropic nature of semiconducting films of discotic liquid crystals, with charge carrier mobilities of up to $2.8\times10^{-3}$cm$^2$/Vs. These properties make discotics an interesting choice for applications such as organic photovoltaics.
\end{abstract}

\maketitle

\section{Introduction}

Discotic liquid crystals (DLC) are molecules with a disc-shape core that is conjugated, surrounded by aliphatic chains to render them liquid-crystalline \cite{chandrasekhar1999}.  The pi-orbitals of the core house delocalized electrons and extend perpendicular to the plane of the disc.  Due to the natural tendency of pi-orbitals of adjacent molecules to maximally overlap, the molecules can self-align into columnar molecular stacks, resulting in the formation of one-dimensional semiconductor wires.  Such a wire is surrounded by entangled aliphatic chains of all the molecules, forming an insulating sheet around the semiconducting wire.  A film of ordered columns is expected to exhibit a highly anisotropic mobility: very high compared to disordered organic materials along the columns and very low perpendicular to the columns.  This is an attractive feature for many applications, as in most semiconductor devices, be it transistors, solar cells or light-emitting diodes, it is the intention to provide electrical conduction in a single direction, and isolation in perpendicular directions \cite{chandrasekhar1999,schmidt-mende2002,boden1999}.

The mobility of discotic liquid semiconducting materials is typically determined using time-of-flight photoconductivity measurements \cite{adam1993,fujikake2004}. Reports on space-charge limited current mobilities of these materials \cite{donley2004}, where also charge carrier injection has to be considered, are scarce. We were not able to find a combination of both complementary methods in literature.

In this letter, we present the realization and characterization of semiconducting devices of aligned discotic liquid crystalline columns between two planparallel electrodes.  Such configuration with the columns perpendicular to the electrode surfaces corresponds to the architecture of relevant organic semiconductor devices such as organic light-emitting diodes and organic solar cells.

\section{Experimental}

	\begin{figure}
		\includegraphics[width=8.5cm]{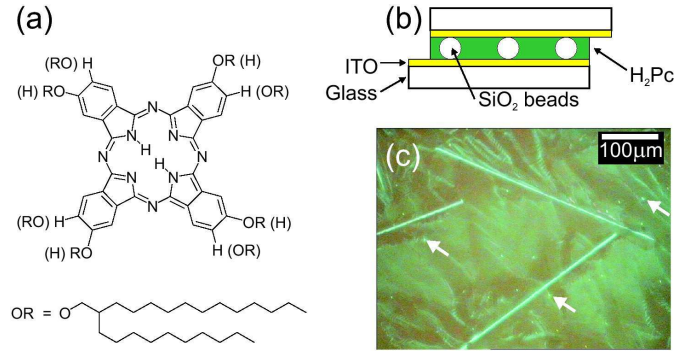}
		\caption{(a) The molecular structure of \htwopc. (b) Symmetric sample cell. (c) Cross-polarized microscopy of a filled sample cell shows homeotropic alignment of the DLC. Three birefringent defect lines and the SiO$_2$ spacers, some of them indicated by arrows, are also visible.}
		\label{fig:1}		
	\end{figure}

The discotic liquid crystal material we used is the discotic  2(3),9(10),16(17),23(24)-Tetra(2-decyltetradecyloxy)-phthalocyanine, abbreviated \htwopc, and shown in Fig.~\ref{fig:1}(a).
The synthesis of the \htwopc molecule is described elsewhere \cite{tant2004phd}. Differential Scanning Calorimetry shows two phase transitions above room temperature: a transition from the liquid crystalline columnar rectangular to columnar hexagonal phase at 333K, and the transition to the isotropic phase in the melt at 453K. A highly ordered liquid-crystalline columnar phase can be attained by heating the material to the isotropic melt and then slowly cooling to room temperature to induce self-alignment of molecular stacks. Further interesting properties of the selected material are that it is processable from solution, and that the peak absorption wavelength is between 600 and 700nm, quite adequate for photovoltaic applications. In order to realize films with homeotropic alignment (i.e., with the column axes normal to the substrate), we proceeded as follows.  Two glass substrates with indium-tin oxide (ITO) contacts of varying width were separated by SiO$_2$ beads, chosen 3 or 5$\mu$m diameter, as shown in Fig.~\ref{fig:1}(b). The phthalocyanine discotic was applied to the edge of this sample cell. Heating to 456K into the isotropic phase lead to capillary action, the liquid \htwopc filled the space between the ITO substrates. The sample was usually cooled at a rate of 1K/min to 426K, then with approx.\ 20K/min back to room temperature. Optical microscopy with crossed polarizers revealed a homeotropic alignment of the discotics between the ITO plates, as seen in Fig.~\ref{fig:1}(c). The micrograph also shows birefringent defect lines typical for homeotropic alignment \cite{vanderpol1989}, and SiO$_2$ spacers. One set of samples was prepared with ITO plates coated with an Octadecyltrichlorosilane (OTS) monolayer by deposition from vapor phase. With this pretreatment, the DLC alignment tends to be homogenous (i.e., column axes parallel to the substrate) and/or random rather than homeotropic, as indicated by optical microscopy. Measurements of conduction perpendicular to the column stacking direction were not only performed on the OTS samples: for verification, we prepared field effect transistor (FET) samples on SiO$_2$ (100nm)/Si/Al substrates with Au bottom-contact. In one set of samples, the DLC was sandwiched between a glass plate and the FET substrate (both not OTS-treated), and aligned homeotropically as described above. As in FET structures the conduction channel is along the substrate, and the DLC was aligned homeotropically, conduction perpendicular to the column stacking direction is looked at, yielding ideally very low currents. We also tried to prepare FET structures for measuring the DLC mobility along the discotic columns by means of different surface treatments, but did not succeed yet. A second set of FET samples was prepared by spin coating the DLC from solution (20mg/ml, solvent: toluene) on top of Au bottom-contact structures.

On the glass/ITO/DLC/ITO/glass samples, we performed frequency-dependent capacitance measurements using an HP 4275 LCR meter, and estimated the device thickness $L$ from the capacitance $C$ at 1 MHz frequency, 
\begin{equation}
	L = \frac{\varepsilon \varepsilon_0}{C} .
	\label{eqn:1}
\end{equation}
Here, $\varepsilon_0$ is the vacuum permittivity. In Eqn.~\ref{eqn:1},  we assumed the equivalence of our samples to a parallel plate capacitor filled with a medium (the discotic) with a dielectric constant $\varepsilon$ of 3. This value is a commonly used estimate for organic materials. For triphenylenes, $\varepsilon=$2.5 was reported \cite{boden1999}. For verification of the device thickness, we checked the SiO$_2$ bead distribution under an optical microscope, which supplies a lower limit.

The temperature-dependent current--voltage measurements as well as the FET measurements were performed using an HP 4156C parameter analyzer with the device on a hotstage. The DLC deposition, and the capacitance and current--voltage measurements were performed in N$_2$ atmosphere, optical microscopy was done in air. In the time-of-flight setup \cite{seynhaeve1989}, pulsed light with a wavelength of 660nm and a pulse width of 6ns was obtained from an LSI nitrogen laser with dye cell attachment. The photocurrent signals were recorded on an Iwatsu 8132 digitizing storage oscilloscope. An HP 214B pulse generator was used as a bias voltage source. The sample was held on a temperature-controlled metal support in a vacuum chamber.

\section{Results and Discussion}

\subsection{Space-Charge Limited Current Measurements}

	\begin{figure}
		\includegraphics[height=7.5cm]{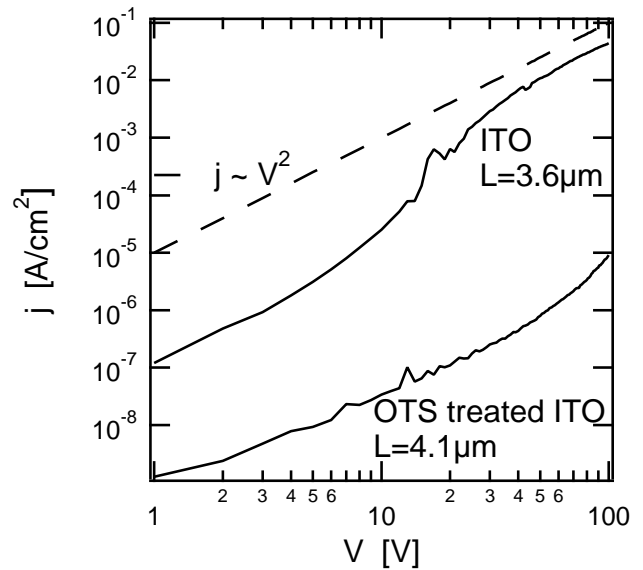}
		\caption{Current density-voltage characteristics of a two sample cells filled with \htwopc, one of them with OTS-treated ITO.}
		\label{fig:2}		
	\end{figure}
	
	\begin{figure}
		\includegraphics[height=7.5cm]{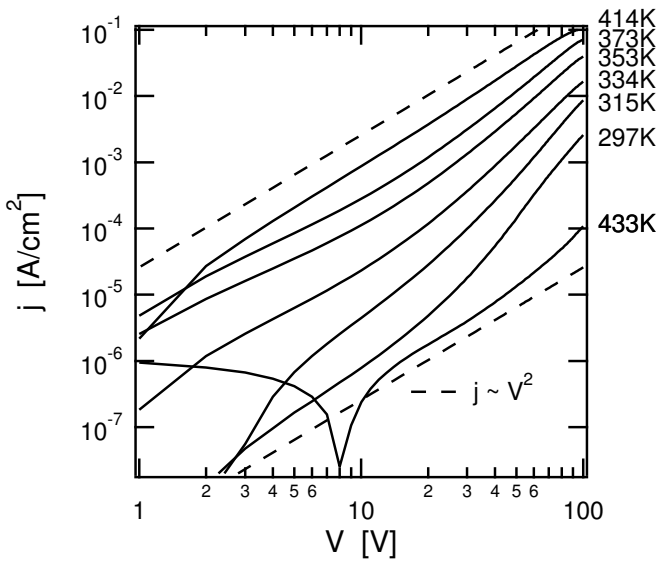}
		\caption{Temperature-dependent current density--voltage characteristics of an \htwopc sample cell (sample thickness 5.0$\mu$m). The shifted zero-crossing of the curve at 433K temperature is due to hysteresis.}
		\label{fig:3}		
	\end{figure}

The space-charge limited current measurements were performed on the glass/ITO/DLC/ITO/glass samples. The effect of homeotropic alignment on the charge transport is demonstrated by the current--voltage measurements shown in Fig.~\ref{fig:2}. A homeotropically aligned \htwopc sample cell is compared to a sample with OTS treated ITO plates, the latter showing a partly homogenous, partly homeotropic alignment by optical microscopy with crossed polarizers. As in this latter sample the charges partly have to be transported perpendicular to the molecular columns, the current density is at least 2 orders of magnitude lower than for the case of completely homeotropic alignment, indicating a high electrical anisotropy. From the current--voltage characteristics, we determined space-charge limited current (SCLC) mobilities in cases where the measurements obey the Mott-Gurney law for trap-free SCLC,
\begin{equation}
	j = \frac{9}{8}\varepsilon\varepsilon_0 \mu \frac{V^2}{L^3} ,
\end{equation}
where $j$ is the current density at the voltage $V$, $\mu$ denotes the drift mobility, $L$ denotes the sample thickness, and $\varepsilon\varepsilon_0$ the electric permittivity of the DLC. The temperature-dependent current--voltage measurements of an \htwopc sample cell (Fig.~\ref{fig:3}) show the characteristic $j \propto V^2$ behaviour. At lower temperatures and intermediate voltages, a slope larger than two indicates an energy-distributed defect state. At higher temperatures, trap-free SCLC can be observed over the whole voltage range measured. The mobilities extracted from SCLC measurements range from $7\times10^{-4}$cm$^2$/Vs at room temperature to about $4\times10^{-3}$cm$^2$/Vs at 414K. The difference of these two values is probably mainly due to the susceptibility of the SCLC method to injection barriers \cite{deboer2004,reynaert2004}. The higher one corresponds to the true SCLC mobilities, the values being comparable to SCLC mobilities measured on CuPc DLC Langmuir-Blodgett films \cite{donley2004}.

\subsection{Field Effect Transistor Measurements}

	\begin{figure}
		\includegraphics[height=7.5cm]{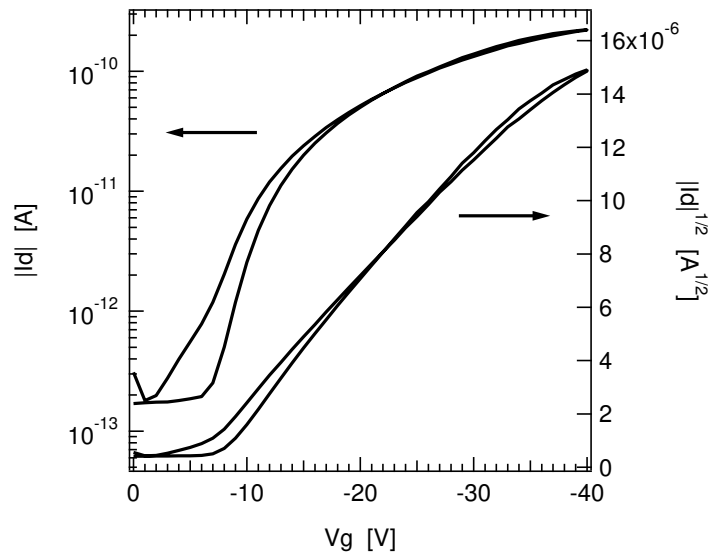}
		\caption{Transfer characteristics in the saturated regime of a disordered \htwopc field effect transistor, spin-coated from solution.}
		\label{fig:4}		
	\end{figure}

As we can neither exclude a possible influence of the OTS monolayer as charge injection barrier, nor possess the means to properly quantify the influence of the OTS treatment on the stacking direction of the discotic columns, we performed measurements on two different types of FET samples for verification (not shown). The first set of samples consists of a homeotropically aligned DLC sandwiched between the FET substrate and a cover glass (channels of 3$\mu$m length and 1mm width).  As the current in FET structures flows from source to drain along the surface of the dielectric layer, we were able to investigate conduction perpendicular to the discotic columns. The resulting source-drain currents were of the order of 10$^{-12}$A for drain and gate voltages of up to $-40$V. No clear gate voltage dependence, and thus no transistor action, could be observed. A second set of FET samples was prepared by spin-coating the DLC from solution on FET substrates, yielding amorphous films (channels of 10$\mu$m length and 1mm width). These structures had very low source-drain currents of the order of $10^{-10}$A, but clearly exhibited typical transistor output and transfer characteristics. The latter are shown in Fig.~\ref{fig:4}. The shape of the FET characteristics indicates that contact resistances are present, but not dominant. From the saturation regime, we determined FET mobilities of about $10^{-7}$cm$^2$/Vs. This low value is mainly due to the disordered nature of the spin-coated film. 

The FET measurements illustrate that charge transport perpendicular to the DLC columns is limited mainly by the anisotropic charge transport properties rather than injection barriers. On basis of these results we also feel justified to rule out a dominant contribution of ionic conduction. The space-charge limited current and FET measurements clearly demonstrate a strong anisotropy of the charge transport properties, in dependence of the stacking direction of the discotics.

\subsection{Transient Photoconductivity}

	\begin{figure}
		\includegraphics[height=7.5cm]{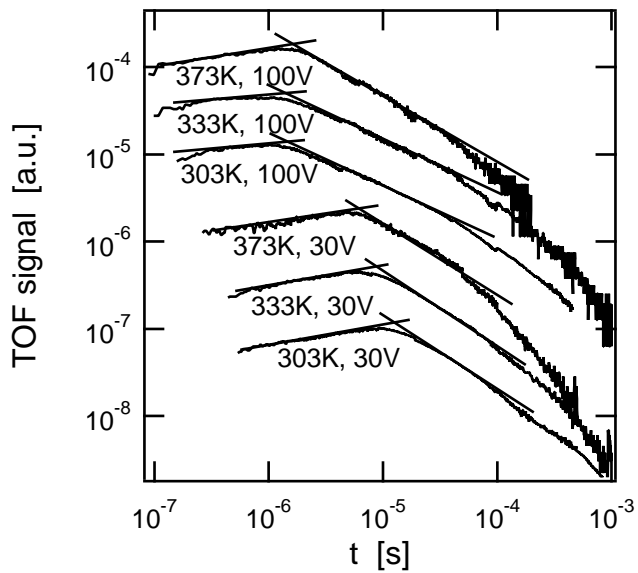}
		\caption{Time-of-flight transient photocurrents of an \htwopc sample cell (sample thickness 6.7$\mu$m). The curves were y-shifted for clarity.}
		\label{fig:5}
	\end{figure}

In the time-of-flight transient photoconductivity measurements, charge carriers were photogenerated by a laser pulse at one electrode of the glass/ITO/DLC/ITO/glass device, and drifted to the other electrode due to an external field. The corresponding displacement current for holes is shown in Fig.~\ref{fig:5}. From the measured photocarrier transit time $\tau$, we determined the mobility$\mu$ using
\begin{equation}
	\mu = \frac{L^2}{V\tau} ,
\end{equation}
where $L$ is the device thickness. For an applied voltage $V$ of 30V, the hole mobilities ranged from $1.0\times10^{-3}$cm$^2$/Vs at 303K to $2.6\times10^{-3}$cm$^2$/Vs at 373K. At 100V, the hole mobilities were between $(2.6-2.8)\times10^{-3}$cm$^2$/Vs. Upon changing the polarity of the external field in order to measure electron mobilities, the transient photocurrents were too low to be detected with our setup. For the \htwopc sample cell with OTS treated ITO electrodes, we also could not measure any photocurrents, supporting the results of the SCLC and FET methods that the mobility perpendicular to the columns is much lower than along them. 

\subsection{Discussion}

The determined values for the hole mobility along the DLC columns using the transient photoconductivity method agree very well with the mobilities we observed using the SCLC method. As the dependence of the mobility determined by these two experimental methods is quite different with respect to voltage and sample thickness, the agreement indicates that the values for the sample thickness calculated using an estimated dielectric constant of 3 are approximately correct. Furthermore, the results support the suggestion that the lower apparent mobilities observed in the SCLC experiments at room-temperature are influenced by contact-limited injection. The \htwopc hole mobilities determined by us correspond well with time-of-flight mobilities measured recently on a CuPc DLC \cite{fujikake2004}. However, in publication~\onlinecite{fujikake2004} electron mobilities of a similar order of magnitude were stated. Similarly, single-crystalline derivatives of phthalocyanines show comparable electron and hole mobilities \cite{cox1974}. As of yet, we were not able to explain this discrepancy. 

We would like to point out that the charge carrier mobilities presented here were determined by experimental techniques in the low carrier concentration regime, such as SCLC and TOF. These measurements are more susceptible to charge trapping as, for instance, FET and pulse-radiolysis time-resolved microwave conductivity measurements \cite{vandecraats1996}, and can therefore yield orders of magnitude lower mobilities.

Our measurements comparing charge transport along and perpendicular to the DLC columns clearly demonstrate a strong anisotropy of the charge transport properties, in dependence of the stacking direction of the discotics.

\section{Conclusions} \label{sec:conclusions}

We have fabricated devices of homeotropically aligned discotic liquid crystal material, and characterized the mobility using time-of-flight photoconductivity and space-charge limited current measurements. The mobility along the columns of discotic molecules attains up to $2.8\times10^{-3}$cm$^2$/Vs in homeotropic alignment and is at least two orders of magnitude lower for transport perpendicular to the columns.

\acknowledgments
This work was performed in the framework of the Soltex project. The authors would like to thank V. Arkhipov for discussions, and S. Put and S. Schols for performing some of the current--voltage measurements.

\bibliography{Papers}

\begin{thebibliography}{13}
\expandafter\ifx\csname natexlab\endcsname\relax\def\natexlab#1{#1}\fi
\expandafter\ifx\csname bibnamefont\endcsname\relax
  \def\bibnamefont#1{#1}\fi
\expandafter\ifx\csname bibfnamefont\endcsname\relax
  \def\bibfnamefont#1{#1}\fi
\expandafter\ifx\csname citenamefont\endcsname\relax
  \def\citenamefont#1{#1}\fi
\expandafter\ifx\csname url\endcsname\relax
  \def\url#1{\texttt{#1}}\fi
\expandafter\ifx\csname urlprefix\endcsname\relax\def\urlprefix{URL }\fi
\providecommand{\bibinfo}[2]{#2}
\providecommand{\eprint}[2][]{\url{#2}}

\bibitem[{\citenamefont{Chandrasekhar and Prasad}(1999)}]{chandrasekhar1999}
\bibinfo{author}{\bibfnamefont{R.}~\bibnamefont{Chandrasekhar}}
  \bibnamefont{and} \bibinfo{author}{\bibfnamefont{S.~K.}
  \bibnamefont{Prasad}}, \bibinfo{journal}{Contemporary {P}hysics}
  \textbf{\bibinfo{volume}{40}}, \bibinfo{pages}{237} (\bibinfo{year}{1999}).

\bibitem[{\citenamefont{Schmidt-Mende et~al.}(2002)\citenamefont{Schmidt-Mende,
  Fechtenk{\"o}tter, M{\"u}llen, Friend, and MacKenzie}}]{schmidt-mende2002}
\bibinfo{author}{\bibfnamefont{L.}~\bibnamefont{Schmidt-Mende}},
  \bibinfo{author}{\bibfnamefont{A.}~\bibnamefont{Fechtenk{\"o}tter}},
  \bibinfo{author}{\bibfnamefont{K.}~\bibnamefont{M{\"u}llen}},
  \bibinfo{author}{\bibfnamefont{R.~H.} \bibnamefont{Friend}},
  \bibnamefont{and} \bibinfo{author}{\bibfnamefont{J.~D.}
  \bibnamefont{MacKenzie}}, \bibinfo{journal}{Physica E}
  \textbf{\bibinfo{volume}{14}}, \bibinfo{pages}{263} (\bibinfo{year}{2002}).

\bibitem[{\citenamefont{Boden et~al.}(1999)\citenamefont{Boden, Bushby,
  Clements, and Movaghar}}]{boden1999}
\bibinfo{author}{\bibfnamefont{N.}~\bibnamefont{Boden}},
  \bibinfo{author}{\bibfnamefont{R.~J.} \bibnamefont{Bushby}},
  \bibinfo{author}{\bibfnamefont{J.}~\bibnamefont{Clements}}, \bibnamefont{and}
  \bibinfo{author}{\bibfnamefont{B.}~\bibnamefont{Movaghar}},
  \bibinfo{journal}{J. Mater. Chem.} \textbf{\bibinfo{volume}{9}},
  \bibinfo{pages}{2081} (\bibinfo{year}{1999}).

\bibitem[{\citenamefont{Adam et~al.}(1993)\citenamefont{Adam, Closs, Frey,
  Funhoff, Haarer, Ringsdorf, Schuhmacher, and Siemensmeyer}}]{adam1993}
\bibinfo{author}{\bibfnamefont{D.}~\bibnamefont{Adam}},
  \bibinfo{author}{\bibfnamefont{F.}~\bibnamefont{Closs}},
  \bibinfo{author}{\bibfnamefont{T.}~\bibnamefont{Frey}},
  \bibinfo{author}{\bibfnamefont{D.}~\bibnamefont{Funhoff}},
  \bibinfo{author}{\bibfnamefont{D.}~\bibnamefont{Haarer}},
  \bibinfo{author}{\bibfnamefont{H.}~\bibnamefont{Ringsdorf}},
  \bibinfo{author}{\bibfnamefont{P.}~\bibnamefont{Schuhmacher}},
  \bibnamefont{and}
  \bibinfo{author}{\bibfnamefont{K.}~\bibnamefont{Siemensmeyer}},
  \bibinfo{journal}{Phys. Rev. Lett.} \textbf{\bibinfo{volume}{70}},
  \bibinfo{pages}{457} (\bibinfo{year}{1993}).

\bibitem[{\citenamefont{Fujikake et~al.}(2004)\citenamefont{Fujikake,
  Sugibayashi, and Ohta}}]{fujikake2004}
\bibinfo{author}{\bibfnamefont{H.}~\bibnamefont{Fujikake}},
  \bibinfo{author}{\bibfnamefont{T.~M.~M.} \bibnamefont{Sugibayashi}},
  \bibnamefont{and} \bibinfo{author}{\bibfnamefont{K.}~\bibnamefont{Ohta}},
  \bibinfo{journal}{Appl. Phys. Lett.} \textbf{\bibinfo{volume}{85}},
  \bibinfo{pages}{3474} (\bibinfo{year}{2004}).

\bibitem[{\citenamefont{Donley et~al.}(2004)\citenamefont{Donley, Zangmeister,
  Xia, Minch, Drager, Cherian, LaRussa, Kippelen, Domercq, Mathine
  et~al.}}]{donley2004}
\bibinfo{author}{\bibfnamefont{C.~L.} \bibnamefont{Donley}},
  \bibinfo{author}{\bibfnamefont{R.~A.~P.} \bibnamefont{Zangmeister}},
  \bibinfo{author}{\bibfnamefont{W.}~\bibnamefont{Xia}},
  \bibinfo{author}{\bibfnamefont{B.}~\bibnamefont{Minch}},
  \bibinfo{author}{\bibfnamefont{A.}~\bibnamefont{Drager}},
  \bibinfo{author}{\bibfnamefont{S.~K.} \bibnamefont{Cherian}},
  \bibinfo{author}{\bibfnamefont{L.}~\bibnamefont{LaRussa}},
  \bibinfo{author}{\bibfnamefont{B.}~\bibnamefont{Kippelen}},
  \bibinfo{author}{\bibfnamefont{B.}~\bibnamefont{Domercq}},
  \bibinfo{author}{\bibfnamefont{D.~L.} \bibnamefont{Mathine}},
  \bibnamefont{et~al.}, \bibinfo{journal}{J. Mater. Res.}
  \textbf{\bibinfo{volume}{19}}, \bibinfo{pages}{2087} (\bibinfo{year}{2004}).

\bibitem[{\citenamefont{Tant}(2004)}]{tant2004phd}
\bibinfo{author}{\bibfnamefont{J.}~\bibnamefont{Tant}}, Ph.D. thesis,
  \bibinfo{school}{Universit{\'e} Libre de Bruxelles} (\bibinfo{year}{2004}).

\bibitem[{\citenamefont{van~der Pol et~al.}(1989)\citenamefont{van~der Pol,
  Neeleman, Zwicker, Nolte, Drenth, Aerts, Visser, and Picken}}]{vanderpol1989}
\bibinfo{author}{\bibfnamefont{J.~F.} \bibnamefont{van~der Pol}},
  \bibinfo{author}{\bibfnamefont{E.}~\bibnamefont{Neeleman}},
  \bibinfo{author}{\bibfnamefont{J.~W.} \bibnamefont{Zwicker}},
  \bibinfo{author}{\bibfnamefont{R.~J.~M.} \bibnamefont{Nolte}},
  \bibinfo{author}{\bibfnamefont{W.}~\bibnamefont{Drenth}},
  \bibinfo{author}{\bibfnamefont{J.}~\bibnamefont{Aerts}},
  \bibinfo{author}{\bibfnamefont{R.}~\bibnamefont{Visser}}, \bibnamefont{and}
  \bibinfo{author}{\bibfnamefont{S.~J.} \bibnamefont{Picken}},
  \bibinfo{journal}{Liq. Cryst.} \textbf{\bibinfo{volume}{6}},
  \bibinfo{pages}{577} (\bibinfo{year}{1989}).

\bibitem[{\citenamefont{Seynhaeve et~al.}(1989)\citenamefont{Seynhaeve,
  Barclay, Adriaenssens, and Marshall}}]{seynhaeve1989}
\bibinfo{author}{\bibfnamefont{G.~F.} \bibnamefont{Seynhaeve}},
  \bibinfo{author}{\bibfnamefont{R.~P.} \bibnamefont{Barclay}},
  \bibinfo{author}{\bibfnamefont{G.~J.} \bibnamefont{Adriaenssens}},
  \bibnamefont{and} \bibinfo{author}{\bibfnamefont{J.~M.}
  \bibnamefont{Marshall}}, \bibinfo{journal}{Phys. Rev. B}
  \textbf{\bibinfo{volume}{39}}, \bibinfo{pages}{10196} (\bibinfo{year}{1989}).

\bibitem[{\citenamefont{de~Boer et~al.}(2004)\citenamefont{de~Boer, Jochemsen,
  Klapwijk, Morpurgo, Niemax, Tripathi, and Pflaum}}]{deboer2004}
\bibinfo{author}{\bibfnamefont{R.~W.~I.} \bibnamefont{de~Boer}},
  \bibinfo{author}{\bibfnamefont{M.}~\bibnamefont{Jochemsen}},
  \bibinfo{author}{\bibfnamefont{T.~M.} \bibnamefont{Klapwijk}},
  \bibinfo{author}{\bibfnamefont{A.~F.} \bibnamefont{Morpurgo}},
  \bibinfo{author}{\bibfnamefont{J.}~\bibnamefont{Niemax}},
  \bibinfo{author}{\bibfnamefont{A.~K.} \bibnamefont{Tripathi}},
  \bibnamefont{and} \bibinfo{author}{\bibfnamefont{J.}~\bibnamefont{Pflaum}},
  \bibinfo{journal}{J. Appl. Phys.} \textbf{\bibinfo{volume}{95}},
  \bibinfo{pages}{1196} (\bibinfo{year}{2004}).

\bibitem[{\citenamefont{Reynaert et~al.}(2004)\citenamefont{Reynaert, Arkhipov,
  Borghs, and Heremans}}]{reynaert2004}
\bibinfo{author}{\bibfnamefont{J.}~\bibnamefont{Reynaert}},
  \bibinfo{author}{\bibfnamefont{V.~I.} \bibnamefont{Arkhipov}},
  \bibinfo{author}{\bibfnamefont{G.}~\bibnamefont{Borghs}}, \bibnamefont{and}
  \bibinfo{author}{\bibfnamefont{P.}~\bibnamefont{Heremans}},
  \bibinfo{journal}{Appl. Phys. Lett.} \textbf{\bibinfo{volume}{85}},
  \bibinfo{pages}{603} (\bibinfo{year}{2004}).

\bibitem[{\citenamefont{Cox and Knight}(1974)}]{cox1974}
\bibinfo{author}{\bibfnamefont{G.~A.} \bibnamefont{Cox}} \bibnamefont{and}
  \bibinfo{author}{\bibfnamefont{P.~C.} \bibnamefont{Knight}},
  \bibinfo{journal}{J. {P}hys. {C}. {S}ol. {S}tate {P}hys.}
  \textbf{\bibinfo{volume}{7}}, \bibinfo{pages}{146} (\bibinfo{year}{1974}).

\bibitem[{\citenamefont{van~de Craats et~al.}(1996)\citenamefont{van~de Craats,
  Warman, de~Haas, Adam, Simmerer, Haarer, and Schuhmacher}}]{vandecraats1996}
\bibinfo{author}{\bibfnamefont{A.~M.} \bibnamefont{van~de Craats}},
  \bibinfo{author}{\bibfnamefont{J.~M.} \bibnamefont{Warman}},
  \bibinfo{author}{\bibfnamefont{M.~P.} \bibnamefont{de~Haas}},
  \bibinfo{author}{\bibfnamefont{D.}~\bibnamefont{Adam}},
  \bibinfo{author}{\bibfnamefont{J.}~\bibnamefont{Simmerer}},
  \bibinfo{author}{\bibfnamefont{D.}~\bibnamefont{Haarer}}, \bibnamefont{and}
  \bibinfo{author}{\bibfnamefont{P.}~\bibnamefont{Schuhmacher}},
  \bibinfo{journal}{Adv. Mater.} \textbf{\bibinfo{volume}{8}},
  \bibinfo{pages}{823} (\bibinfo{year}{1996}).

\end{thebibliography}
\bibliographystyle{apsrev}

\end{document}